# Observation of the Anomalously Slow (Power-Law) Relaxation of the System of Interacting Liquid Nanoclusters in the Disordered Confinement of a Random Porous Medium

V. D. Borman, A. A. Belogorlov, and V. N. Tronin

Department of Molecular Physics, National Research Nuclear University MEPhI, Kashirskoe sh. 31,

Moscow, 115409 Russia

## Abstract

The time evolution of the system of water in the Libersorb 23 (L23) disordered nanoporous medium after the complete filling at excess pressure and the subsequent removal of excess pressure has been studied. It has been found that three stages can be identified in the relaxation of the L23–water system under study. At the first stage, a portion of water at the removal of excess pressure rapidly flows out in the pressure reduction time, i.e., following a decrease in the pressure. It has been shown that, at temperatures below the dispersion transition temperature $T < T_d = 284$ K, e.g., $T = 277$ K, the degree of filling $\theta$ decreases from 1 to 0.8 in 10 s, following the variation of excess pressure. At the second stage of relaxation, the degree of filling $\theta$ varies slowly according to a power law $\theta \sim t^{-\alpha}$ with the exponent $\alpha < 0.1$ in the time $t \sim 10^5$ s. This corresponds to a slow relaxation of the formed metastable state of the nonwetting liquid in the porous medium. At the third stage when $t > 10^5$ s, the formed metastable state decays, which is manifested in the transition to a power-law dependence $\theta(t)$ with a larger exponent.

The extrusion-time distribution function of pores has been calculated along with the time dependence of the degree of filling, which qualitatively describes the observed anomalously slow relaxation and crossover of the transition to the stage of decay with a power-law dependence $\theta(t)$ with a larger exponent.

It has been shown that the relaxation and decay of the metastable state of the confined nonwetting liquid at $\theta > \theta_c$ are attributed to the appearance of local configurations of liquid clusters in confinement and their interaction inside the infinite percolation cluster of filled pores.

## 1. INTRODUCTION

The states and properties of disordered media such as glasses, colloids, polymers, and loose materials have been actively studied in recent years [1–12]. Numerical investigations are performed [6, 7, 16, 18], and phenomenological models such as shear transformation zone (STZ) [1, 4, 9, 13, 19], dynamic heterogeneity (DH) [2, 5, 21], random first order transition theory (RFOT), topological bond-oriented local configuration [1, 6, 7, 16, 18, 27, 28], etc. (see, e.g., [1–3, 10, 14, 20]) were introduced and discussed. These models involve the concept of local structures (configurations) determining the properties of disordered media and are used to describe states and relaxation of glasses, colloids, polymers, and loose media, as well as liquid–glass transitions and the sol–gel process resulting in the appearance of random order. According to [1, 22, 23], these media are nonergodic and are characterized by an anomalously slow relaxation of local nonequilibrium states, which is phenomenologically described by a stretched exponential relaxation law [1, 8, 12, 27]. Anomalously slow relaxation means that the system under study cannot reach any point of the phase space in any large relaxation observation time smaller than the lifetime of the states [1]. Anomalously slow (power-law) relaxation is attributed in phenomenological models of disordered media to the formation and decay of metastable states of random local structures in a disordered medium.

Disordered media include disordered porous media. Their structure is studied by gas adsorption–desorption and mercury (liquid) porometry methods. These methods provide information on the volume of pores, area of their surface, and pore size distribution function. New possibilities for investigating the properties of disordered porous media are opened by studying the states of a nonwetting liquid confined in the disordered structure of pores after the removal of excess pressure [24–26].

The aim of this work is to study the detected anomalously slow relaxation of metastable states of the liquid confined in a porous medium. We studied the porous medium that is a silica gel where the disordered structure of the frame and pores appears in a sol–gel process. The process of relaxation is the anomalously slow extrusion of the confined nonwetting liquid from the porous medium. The relaxation of such states has not yet been studied, but the anomalously slow extrusion of the nonwetting liquid from the porous medium was sometimes observed (see, e.g., [29–32]). An approach to the description of anomalously slow relaxation state of the confined nonwetting liquid is proposed with allowance for the appearance of local configurations of liquid clusters in confinement, their energetics, and interaction inside the infinite percolation cluster of filled pores. Such an approach makes it possible to qualitatively explain observed power-law relaxation.

The porous medium immersed in a nonwetting liquid can be filled with this liquid only when excess pressure is above a certain critical value, which can be estimated by the Laplace formula. After the subsequent removal of excess pressure, such a system can be in an unstable state, because surface forces push the nonwetting liquid out. These phenomena are identified in numerous studies of the intrusion–extrusion hysteresis of the nonwetting liquid from the porous medium immeresed in this liquid [31, 33–43]. The characteristic liquid extrusion time was determined in the experiments on the fast compression of the porous medium immersed in the liquid and the subsequent fast removal of excess pressure [44]. For the hydrophobized L23 silica gel with the mean radius of pores $\bar{R} \sim 4\,nm$ and the size of granules 10 µm and water, the complete liquid extrusion time was ~0.1 s. The fast extrusion mechanism was proposed in [44] within the dynamic percolation theory [44, 45].

At the same time, it is known that the confinement of a fraction of the liquid or the entire liquid after the removal of excess pressure is observed for many nonwetting liquids and disordered porous media. In this case, fast extrusion can be observed only for a fraction of the liquid. The confinement of the liquid was detected when studying intrusion–extrusion for water, aqueous solutions of salts and organic substances, mercury and other metals, Wood's alloy, hydrophobized silica gels KS K-G, PEP 100, PEP 300, Fluka 60, Fluka 100, C8W (Waters), and Vycor and CPG porous glasses [31–43, 48, 49]. These media differ in the degree of hydrophobicity, porosity, average pore size in the range of 0.5–20 nm, and average granule size in the range of 10–100 μm, and the width of the pore size distribution. We note that the confinement effect is not related to a change in the phase state of the liquid. Indeed, according to [29, 50], at the radius of pores $R > 1$ nm and $T > 273$ K, the properties of the liquid in confinement do not differ from the properties of the bulk liquid. The confined liquid can remain in a porous medium for hours, days, and months. For the (mercury–porous glass) system, the weight of porous glass samples with confined mercury and the results of neutron scattering experiments did not change for several months [29]. For other systems consisting of mercury and Vycor and CPG porous glasses and silica gels, it was found that the volume of the confined liquid depends on the extrusion observation time and the size of granules of the porous medium [51].

Investigations indicate that the volume fraction of the confined liquid θ can be from several to hundred percent [31–33, 43, 46, 49, 52, 53]. In particular, for the L23–water system at $T = 279$ K, confinement of ~ 100% liquid changes at the variation of the temperature by $\Delta T \approx 10$ K to a small, ~5 %, volume fraction of the confined liquid. If the degree of filling $\theta$ of pores with the confined liquid is above the percolation threshold $\theta_c$ ($\theta_c = 0.15 - 0.35$, depending on the model of the porous medium [54]) through filled pores, the confined liquid can remain in the porous medium for an observation time of 10–$10^5$ s although the liquid can be extruded through filled pores, which form the «infinite» percolation cluster at $\theta > \theta_c$. If θ is below the percolation threshold, $\theta < \theta_c$, only isolated clusters of filled pores are formed in the porous medium. Paths for extrusion of the liquid from these clusters through filled pores are absent. A possible mechanism of the extrusion of the liquid from clusters of filled pores can be recondensation, i.e., capillary evaporation and subsequent capillary condensation at the (porous medium–surrounding liquid) interface. Capillary condensation and capillary evaporation constitute a mechanism considered as responsible for transport in the experiments on the adsorption–desorption of gases in porous media [55, 56]. In such experiments, a porous medium is placed in a gas atmosphere and the mass of the adsorbed gas is measured as a function of the pressure of the gas near the pressure of the saturated vapor. It is noteworthy that, in view of a low density of the gas as compared to the density of the liquid ($\rho_g \sim (10^{-4} - 10^{-3})\rho_l$), under equal other conditions, the flux of the gas is much smaller and the characteristic adsorption–desorption time in such experiments is much larger than the respective values in the case of the transport of the liquid through liquid-filled pores forming the infinite percolation cluster. This characteristic time in the experiments on the adsorption–desorption of gaseous nitrogen at $T = 77$ K can be ~$10^4$–$10^5$ s, and the characteristic time at the intrusion–extrusion of the nonconfined fraction of the liquid is ~ 0.1 s [49].

Thus, it can be expected that there are two different scenarios of relaxation after the filling of the porous medium and the subsequent removal of excess pressure. In the first scenario, the most part of the liquid is in an unstable state and can be extruded from the porous medium in a characteristic hydrodynamic time of ~ 0.1 s. The remaining part with the fraction of filled pores below the percolation threshold $\theta < \theta_c$ forms clusters of filled pores surrounded by empty pores. The characteristic time of the extrusion of the liquid from these pore clusters should be determined by the process of capillary evaporation of the liquid from the surface of menisci in filled pores and capillary condensation on the surface of menisci with nearly identical curvature at the interface between granules of the porous medium and the bulk liquid. Consequently, the mass flow in this process is expected to be much smaller and the time is expected to be much larger than the respective values in the case of a hydrodynamic flow of the liquid through filled pores. In the second scenario, a fraction of the liquid after intrusion at excess pressure and removal of excess pressure can be in an unstable state and can be extruded in a hydrodynamic time of ~ 0.1–1 s through filled pores of theinfinite percolation cluster. In this work, we study the process of extrusion of the confined liquid at the degree of filling $\theta$ above the percolation threshold $\theta_c$. Relaxation at $\theta < \theta_c$ can be described within the evaporation–condensation mechanism considered in [43].

In [31, 32], it was found that the volume of the confined liquid in the systems consisting of water and theL23 or Fluka 100 C18 porous medium depends critically both on the initial degree of filling and the temperature (dispersion transition). These properties cannot be explained within the concept of intrusion–extrusion of the liquid from individual pores based on the Laplace relation and phenomenological contact angle [46]. These dependences mean that the confinement of a fraction of the nonwetting liquid can be attributed to the collective interaction between liquid clusters in neighboring pores depending on the degree of filling. Correlation effects of the interaction between liquid clusters in confinement were considered in [57–59]. The lattice gas model was used in [58, 59] to describe the state and relaxation of an ensemble of liquid clusters in pores. Such a model allows the inclusion of the interaction between clusters with allowance for the random distribution of neighboring pores in the lattice. Transport in such a model is considered as the diffusion transport of a vapor from a filled pore to a neighboring empty pore. This approach describes states and relaxation at the adsorption–desorption of the gas in terms of capillary condensation and capillary evaporation in the porous medium immersed in a gas atmosphere [60]. The slowing down of the process of desorption in the hysteresis region observed in [61] for the system of cyclohexane in the Vycor porous medium is explained within the lattice gas model by the slowing down of the diffusion of the vapor in the system of pores partially filled as a result of the fragmentation of the condensed liquid. The numerical Monte Carlo study of the lattice gas model with the Glauber–Kawasaki algorithm under the assumption of the diffusion transport of the vapor through neighboring empty pores showed that the volume of the remaining liquid increases with a decrease in the observation time [51, 58, 59].

The confinement of the nonwetting liquid in the disordered structure of pores and the kinetics of the dispersion transition were described in [62] within the analytical percolation theory and statistical theory of fluctuations. Within such an approach, the extrusion of the liquid and its confinement was described for the ground state of the disordered porous medium, which is characterized by the formation of the infinite fractal percolation cluster of filled pores. The observed confinement of the liquid is explained by the transition of a fraction of the liquid filling the medium at excess pressure and the subsequent removal of excess pressure to a metastable state. The energy barrier of the metastable state is determined as the difference between the surface energy of the interaction of a liquid cluster in a pore with the frame of the medium, which «extrudes» the nonwetting liquid, and the surface energy of the «multiparticle interaction» of the liquid cluster in the pore with liquid clusters in neighboring pores. The energy of this «multiparticle interaction» is the difference between the surface energies of liquid clusters in local configurations of empty and filled neighboring pores and the same liquid clusters in isolated pores. The difference between the surface energies of the liquid clusters is due to the disappearance of menisci of the liquid in the mouths of the throats connecting neighboring pores. This means that the surface energy of two clusters in neighboring pores is lower than the surface energy of two isolated clusters. A decrease in the surface energy can be treated as the effective attraction between clusters. With an increase in the concentration of liquid clusters (the degree of filling) and, thereby, with an increase in the number of filled neighboring pores, the energy of «multiparticle attraction» can become higher than the energy of the interface between the liquid and the frame of the porous medium, which extrudes the nonwetting liquid. In this case, the energy barrier of the metastable state is positive and the extrusion becomes energetically unfavorable for some pores. For the other filled pores, this barrier is negative and the liquid spontaneously flows from these pores in the hydrodynamic flow time through filled pores of the percolation cluster.

For the disordered porous medium partially filled with the liquid with the pore size distribution, the mutual arrangement of filled and empty pores is random both on the fractal shell of the percolation cluster of filled pores and inside the percolation cluster. Consequently, the energy barrier of the metastable state forms a random potential profile in the space of the porous medium on the shell and in the volume of the percolation cluster of filled pores. The extrusion of the liquid from pores is the process of overcoming of a set of random maxima of the potential profile. In the process of extrusion of the liquid, the degree of filling of pores decreases and the percolation cluster «is contracted». This can accelerate the extrusion of the confined liquid because of a decrease in θ (a decrease in the number of neighboring filled pores) and a decrease in the energy of the «multiparticle interaction». As a result, the barrier of metastable states is reduced and the extrusion rate increases.

The description of the relaxation of the metastable state of the confined liquid based on the calculation of the distribution function of filled pores in liquid extrusion time was proposed in [62]. This description of relaxation is used in this work for the porous medium with a narrow pore size distribution with the relative width $\Delta R/R << 1$.

The experiments described below were performed in order to test the mechanism proposed in [62] for the relaxation of the state of the liquid confined in the porous medium immersed in this liquid at degrees of filling with the confined liquid above the percolation threshold. The problem was formulated as follows. We chose the system of water in the Libersorb 23 (L23) disordered nanoporous medium. After the complete filling at excess pressure and the removal of excess pressure in this system, two states of the confined liquid are observed at temperatures above and below the dispersion transition temperature $T_d \approx$ 284 K. In one of them, the degree of filling with the confined liquid $\theta$ at $T < T_d = 284$ K is higher than the percolation threshold $\theta_c$ and the liquid can be extruded. In the other state, the degree of filling with the confined liquid at $T > T_d = 284$ K is $\theta < \theta_c$, isolated clusters of filled pores are in the metastable state and the extrusion of the liquid is possible through the mechanism of capillary evaporation–condensation at the boundary of a granule of the porous medium immersed in this liquid.

When measuring the time dependence $\theta(t)$, it was necessary to avoid the systematic error at the continuous determination of $\theta$ for a long time up to $10^5$ s. We developed and used the method of the step-by-step determination of $\theta$ at different times (Section 2). This method made it possible to reproduce and to contr the initial state of the confined liquid at times $t > 10$ s larger than the time of spontaneous, barrierless extrusion of a fraction of the liquid at $\theta > \theta_c$.

We detected an anomalously slow extrusion of the liquid at the initial degrees of filling with the confined liquid $\theta$ both above and below the percolation threshold (Subsection 2.2). In the observation time of this slow relaxation, the experimental data at $\theta > \theta_c$ are described by an inverse power dependence with the exponent $\alpha < 0.1$. It was found that the exponent $\alpha$ depends on the temperature, reaching a maximum near the dispersion transition temperature $T_d = 284$ K. This can indicate a change in the transport mechanism.

The aim of the theoretical analysis of the results was to describe the general picture of the appearance and relaxation of the metastable state of the nonwetting liquid in the disordered porous medium. We propose a new relaxation mechanism determined by the interaction between local nonequilibrium configurations in the «infinite» percolation cluster and describe for the first time stages of the formation, relaxation, and decay of the metastable state. The time dependences of the volume of the confined liquid for the case $\theta > \theta_c$, where the liquid can flow through filled pores of the «infinite» percolation cluster are obtained for the first time.

## 2. EXPERIMENT

### 2.1. *Porous Medium*

The experiments were performed with the Libersorb 23 (*L*23) nanoporous medium. It is the KSK-G silica gel, where the disordered structure of pores is formed in the sol--gel process. The surface of pores of the KSK-G silica gel was modified by alkylsilanes at the laboratory headed by Prof. G.V. Lisichkin (Moscow State University) in order to ensure hydrophobic properties [63, 64]. The characteristics of the samples of the *L*23 porous medium were obtained from the adsorption of nitrogen at an Autosorb IQ (Quantachrome, USA) analyzer for studying low-temperature sorption and a Micro-Ultrapyc 1200e helium pycnometer. The density of the *L*23 porous medium was $\rho = (1.7798 \pm 0.0016)$ g/cm$^3$, the specific volume of pores was $V_p = (0.62 \pm 0.02)$ cm$^3$/g, the porosity of the material was $\varphi = 0.52$, the specific surface was $S_p = (199 \pm 7)$ m$^2$/g, and the mean size of granules of the *L*23 powder was ~ 10 μm. The pore volume distribution function according to BJH was shown in Fig. 1. This function provides only qualitative representation of the pore radius distirbution. The mean radius of pores is $<R> = (5.0 \pm 0.2)$ nm, the FWHM of the distribution near the maximum is $\Delta R = (0.4 \pm 0.1)$ nm, and $\Delta R/<R> \leq 0.1$. Tails of the distribution are also observed in the regions of large and small sizes of pores. Consequently, since $\Delta R/<R> << 1$, the L23 porous medium can be used to test the relaxation model [62], where the random potential field of the barrier of local metastable states of the confined liquid (water) is attributed to the fractality of the «infinite» percolation cluster of pores filled with the trapped liquid and a power law of relaxation of such states should be observed.

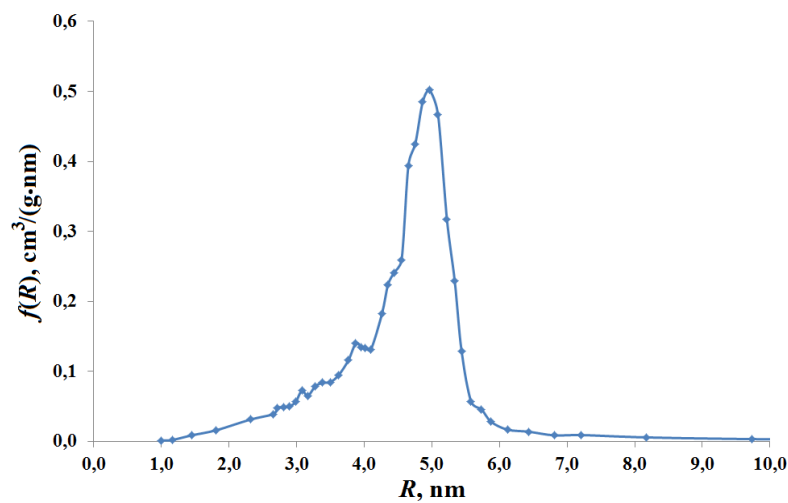

Fig. 1. Pore size distribution function for the Libersorb 23 porous medium according to the BJH method.

### 2.2. *Measurement Procedure*

The aim of the performed measurements was to determine the time dependence of the volume fraction of pores θ filled with the confined liquid. The confined liquid is defined as the fraction of the liquid that remains in pores after the complete filling of pores at an increase in the pressure and the subsequent removal of excess pressure. The other fraction of the liquid in pores flows from pores rapidly in the pressure decreasing time. The extrusion time of this fraction of the liquid from the L23–water system is smaller than 1 s [57]. The time dependence of the volume fraction of pores θ with the confined liquid in a slow liquid extrusion process, i.e., the relaxation of the state of the liquid confined in the porous medium, was studied in this work. The scale of the characteristic liquid extrusion time can differ for the mechanism of liquid transport through the system of filled pores and the mechanism involving capillary evaporation and subsequent capillary condensation at the interface between the bulk liquid and a granule of the powder of the porous medium immersed in this liquid. The latter mechanism can obviously be efficient in the case of the formation of an «infinite» percolation cluster of empty pores, i.e., when the degree of filling with the confined liquid is $\theta < \theta_{c1} = 0.7$–$0.85$. Therefore, the developed method should ensure the measurement of the volume at the extrusion and evaporation of the confined liquid. In experiments, it was also necessary to determine the initial state of the confined liquid, the time and conditions of the beginning of relaxation of the state of and of the extrusion of the confined liquid from the porous medium, and the possible effect of the prehistory of the formation of the initial state. A particular goal is to experimentally confirm the conservation of random properties of the medium with the confined liquid in a long-term experiment at repeated measurements. The determination of the time scales of extrusion and evaporation of the confined liquid makes it possible to determine the time interval in which, according to the model [62], the slow extrusion of the liquid should occur, according to the predicted power law of relaxation.

The method used to measure the volume of the (nonwetting liquid–porous medium) system was similar to mercury [65] or water [63] porometry. A preliminarily dried and degased sample of the *L*23 porous medium with the mass up to 6 g in a container permeable for water was placed in a high-pressure chamber. The remaining free volume of the chamber (28 cm$^3$) was completely filled with distiller water. A rod was introduced into the chamber through seals. The chamber was equipped with a thermostating system, which allowed studies in the temperature range from 243 to 393 K. Before the measurements, the prepared chamber was aged at a given temperature for no less than 1 h. The temperature in the process of measurements was maintained with an accuracy of ±0.2 K. The chamber was mounted on the bench described in [31]. The bench ensured the programmed force action on the rod of the chamber, measurement of the force, and displacement of the rod. Pores of the porous medium were filled under pressure ensuring by the compression of the system in the closed volume at the introduction of the rod into the chamber. The impermeability of the chamber filled with water and porous medium was ensured by seals. The force (*F*) acting on the rod was measured by a strain gauge dynamometer CWH-T2 (Dacell,

South Korea), and the displacements ($l$) of the rod were measured by a potentiometric displacement sensor model 8719 (Burster, Germany). The pressure ($p$) in the chamber was determined as $p = F/S_r$, where $S_r = 0.785$ cm$^2$ is the area of the cross section of the rod. A change in the internal volume of the chamber ($V$) was determined as $V = l \cdot S_r$. Data were recorded from gauges with a frequency of 1 kHz.

An increase in the pressure in the chamber resulted in the elastic strains of the chamber, liquid, and porous medium. Filling of pores of the $L23$ porous medium with water was observed only at pressure > $120 \cdot 10^5$ Pa. This made it possible to determine the effective compressibility of the chamber, as well as the compressibilities of water and porous medium in additional experiments at pressure < $120 \cdot 10^5$ Pa; the corresponding data were taken into account when determining the volume of filled pores at a given pressure.

In the experiments, we used the step-by-step method to determine the time ($t$) dependence of the degree of filling ($\theta$) of pores with the confined liquid. The degree $\theta$ was determined after the observation time ($t_i$) of the liquid flowing out via extrusion or evaporation. Then, the (porous medium–liquid) system was returned to the initial state, and $\theta(t_j)$ was determined at different times $t_j$, both larger and smaller than $t_i$. The dependence $\theta(t)$ was obtained by multiple repetition of such a procedure with various $t_i$ values. Each step of the procedure consisted of two successive intrusion–extrusion cycles performed after the time $t_i$. The pressure dependences of a change in the volume of the $L23$–water system at $T = 286$ K for two successive cycles at the first step of the measurement are shown in Fig. 2. The dependences in Figs. 2a and 2c for the preliminarily dried and degased porous medium are presented after the subtraction of the change in the volume of the system because of the compressibility of the liquid, the frame of the porous medium, and the chamber. Dependence I corresponds to a decrease in the volume of the empty porous medium at an increase in the pressure in the first cycle. Dependence I' corresponds to an increase in the volume at a decrease in the pressure. Three sections can be identified in the dependence $V(p)$ at an increase in the pressure. Section 0–1 corresponds to the compression of the sample of the empty porous medium, and the slope of this linear section is determined by the compressibility of the empty porous medium. It is noteworthy that the filling of pores of granules of the porous medium immersed in the liquid in the chamber was not observed both in the initial state at atmospheric pressure $p = (1.00 \pm 0.05)$ $\cdot 10^5$ Pa, when empty pores can contain saturated water vapor, and at high pressures up to $120 \cdot 10^5$ Pa at point 1 in section 0–1 when the observation time is smaller than 15 h. Section 1–2 of dependence I corresponds to filling of available pores of the empty porous medium. At pressure $p = 450 \cdot 10^5$ Pa, $\approx$ 99.8% pores are filled. The volume of all filled pores of the sample is equal to the difference $|V_2 - V_1|$. When the pressure decreases below point 3 in dependence I', $V$ decreases, corresponding to the extrusion of the liquid from pores, which is described by dependence I'. The main decrease in the quantity $V$ is observed at low pressures comparable with the error of the measurement of the pressure, and the volume of the confined liquid can be determined from the intersection of the dependence $V(p)$ with the $y$ axis only with the accuracy $\delta V / V \sim 1$. For this reason, in the experiments, we determined the volume of pores in two successive intrusion–extrusion cycles.

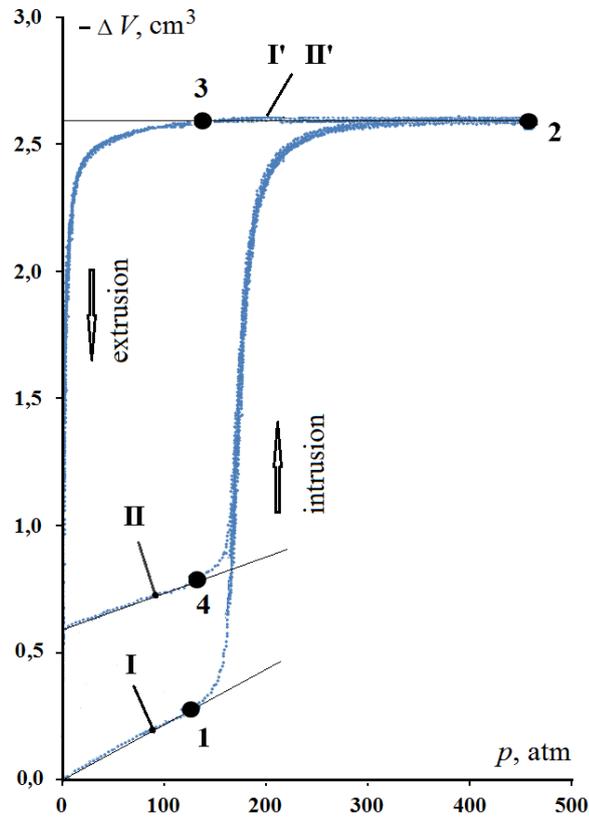

Fig. 2. Pressure dependences of a change in the volume of the $L23$–water system at $T = 286$ K for two successive cycles.

Before repeated filling in the second cycle, some pores contain the confined liquid and, when the pressure in the second cycle increases above the point 4 in dependence II, only a fraction of empty pores are filled. The volume of these empty pores is $|V_2 - V_4|$, and the difference between the volumes of filled pores in the first and second cycles determines the volume of pores filled with the liquid confined after the first cycle, $V_t = (|V_2 - V_1| - |V_2 - V_4|)$. Correspondingly, $\theta = V_t /|V_2 - V_1|$. The relative compressibilities in the sections of filling of the empty porous medium and partially filled porous medium are almost the same, $7.5 \pm 0.9 \ 10^{-9} \ \mathrm{Pa}^{-1}$, within the error of measurements. After the complete filling, the system «forgets» the prehistory of the formation of the preceding initial state and its relaxation, and the initial state of the confined liquid is formed again after the extrusion of a fraction of the liquid (dependence II'). Thus, the volume of all empty pores can be determined in the first filling. The volume of pores that remain empty after the extrusion of the liquid in the time interval from the beginning of extrusion in the first cycle (point 3) to the time of the beginning of filling at the repeated increase in the pressure (point 4) is determined at repeated filling.

Consequently, $V$ in such a method is determined for the time $t$ passing from the beginning of the extrusion of the liquid in the first cycle when the pressure decreases below point 3. The time $t$ includes the time $t_1$ of reaching «zero» excess pressure (i.e., $p = p_{atm} = (1.00 \pm 0.05) \cdot 10^5$ Pa) and the time $t_i$ to the beginning of the repeated filling at point 4 in the second cycle. According to [62], spontaneous barrierless extrusion of the unconfined liquid from some pores, as well as extrusion from pores in which the liquid is in a loosely bound state with a low barrier comparable to the temperature, can occur in the time $t_1$. The height of this barrier can be estimated assuming that $t \sim \tau_0 e^{\frac{E}{T}}$. At $t_1 \sim 10$ s and $\tau_0 \sim 0.1$ s, $E/T \leq 5$. As follows from the below estimate (see section «Discussion of the results»), the volume fraction of pores with the liquid in loosely bound states for the system under study is several percent. Since the pressure decreasing time $t_1 = 10$ s is much larger than the time of hydrodynamic extrusion of a fraction of the liquid at $\theta > \theta_c$, the decrease in the volume adiabatically follows the decrease in the pressure. Thus, the initial state of the system is formed in the time fixed in all measurements as $t_1 = 10$ s, and the relaxation of the metastable state of the confined liquid begins after the time $t_1$. The volume of the confined liquid in the metastable state $V_t$ in the experiments is referred to the observation time of relaxation (extrusion of the liquid) $t = t_i$. This volume includes the volume $V$ of the liquid remaining after the extrusion in the time interval $t = t_i$ and does not include the volume of the liquid in these loosely bound states.

In order to test the reproducibility of the initial state, we performed additional experiments, where the step-by-step method for the determination of $\theta(t)$ was supplemented with multiple measurements of $\theta$ at the chosen time $t = t_1 + t_i = 1$ min. It was found that the spread of the measured $\theta$ values at $t = 1$ min is smaller than the error of the measurements. Other additional experiments indicated that $\theta(t_i)$ values and the $\theta$ value at the different time $t_j$ are independent of the sequence of these measurements. Multiple intrusion–extrusion–confinement experiments with the nonwetting liquid at temperatures $T = 277$–293 K including the subsequent removal of the confined water from the $L23$ porous medium by evacuation drying at $T = 343$ K performed before and after the measurements of $\theta(t)$ confirm that the properties of the porous medium did not change. The specific volume of pores, the pressure of beginning of filling $p_{in}$ at point 1 in Fig. 2, and the pressure of beginning of extrusion of the liquid $p_{out}$ at point 3 remained unchanged within the error of measurements after more than 100 identical cycles. This confirms the stability of the hydrophobizing layer on the surface of the L23 porous medium.

## 2.3. *Experimental Results*

Figure 3 shows the experimental values of the degree of filling $\theta$ of pores with the confined liquid at different observation times $t$ of relaxation for temperatures $T = 279$ and 293 K. These experimental values of $\theta$ were obtained by the step-by-step method in two intrusion–extrusion cycles with the variation of the observation time $t$ of relaxation from 10 to $10^5$ s at each step. Points in Fig. 3 represent the results up to six measurements of θ at each value of the time $t$. The spread of θ values does not exceed the error of the measurement. This confirms the reproducibility of the initial state of the (liquid–disordered porous medium) system at «losing memory» of the preceding states after complete filling in the first and second cycles. The coincidence of θ values can be attributed to the fact that, after complete filling and subsequent removal of excess pressure, the system was in one of the numerous degenerate states with the formation of the «infinite» percolation cluster of filled pores at θ > θ$_c$ or empty pores at θ < θ$_{c1}$. The degree of filling is the same for all these states.

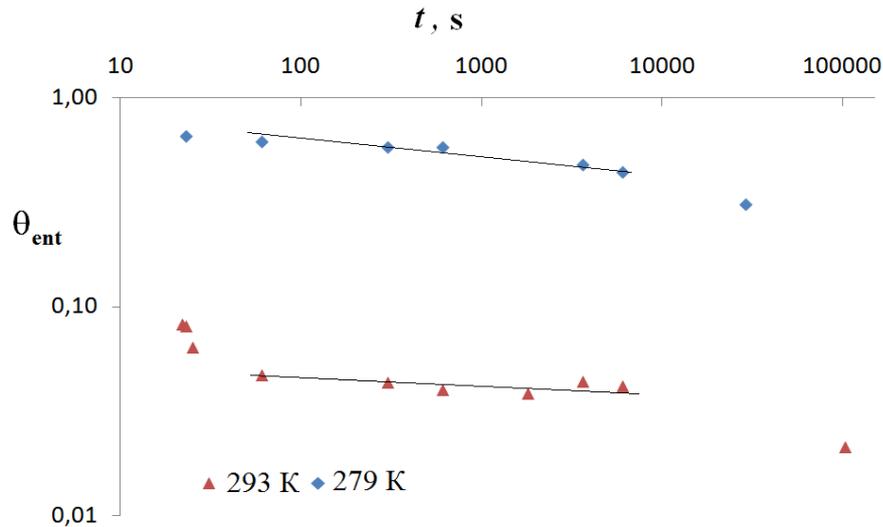

Fig. 3. Values θ at different observation times $t$ of relaxation for temperatures $T$ = 279 and 289 K.

The (water–L23 porous medium) system between the temperatures $T$ = 293 and 279 K undergoes a dispersion transition, when the liquid dispersed in the porous medium at a decrease in the temperature is in a metastable state [24]. With an increase in the temperature, the transition from an almost total extrusion of the liquid at $T$ = 279 K to an almost total extrusion at $T$ = 293 K occurs adiabatically in a narrow temperature range ($\Delta T \approx$ 10 K) near the critical temperature. At the temperature $T$ = 279 K and the observation time of relaxation from 13 to $3 \cdot 10^5$ s, the degree of filling decreases from 0.8 to 0.6 (see Fig. 3). In this case, θ > θ$_c$, where θ$_c$ = 0.15–0.3 is the percolation threshold, and time relaxation occurs under the conditions of the formation of the «infinite» percolation cluster of filled pores. Consequently, the liquid can leave the porous medium via the mechanism of extrusion through filled pores of the percolation cluster. At the temperature $T$ = 293 K in the time interval from 13 to $10^5$ s, the degree of filling θ$_{ent}$ of pores with the liquid varies from 0.08 to 0.02. At such degrees of filling below the percolation threshold θ$_c$, the confined liquid is located in pores that form clusters surrounding by empty pores. Therefore, the volume of the confined liquid can decrease via the gas (water) transport mechanism through capillary evaporation from the surface of the liquid in clusters of filled pores, diffusion of vapor in the residual gas in the infinite percolation cluster of empty pores, and capillary condensation at the interface between the bulk liquid and granules of the porous medium. Under the same other conditions, fluxes in such a transport mechanism are small as compared to the hydrodynamic flow of the liquid because the density of vapor is much lower (by a factor of $10^3$–$10^4$). This can moderate the extrusion of the liquid from individual clusters of filled pores at $T = 293$ K as compared to the faster transport at $T = 279$ K. Such a picture is observed at the temperature $T = 293$ K (see Fig. 3) at small times $t = (13–60)$ s of the extrusion of the liquid from metastable states. A comparatively slow extrusion of the liquid from clusters of filled pores is also indicated by the observed decrease in θ at $T = 293$ K only at the times $t > 10^4$ s. This makes it possible to assume that the experimental data on θ(t) at $T = 279$ K and at θ > θc correspond to the mechanism of hydrodynamic transport of the liquid through filled pores of the percolation cluster in the time interval $10$ s $< t < 6·10^3$ s.

According to Fig. 3, slow relaxation of θ(t) for $T = 279$ K was observed in the time interval $60$ s $< t < 6·10^3$ s, and the log–log plot of the dependence θ(t) (Fig. 4) is described by a straight line within the error of the measurements. A similar picture is observed at $T = 293$ K for the different, gas transport mechanism in the time interval $60$ s $< t < 6·10^3$ s. However, the slope of the dependence θ(t) at this higher temperature is smaller and, therefore, the relaxation rate is lower according to a slower gas transport at $T = 293$ K.

The dependences θ(t) for seven temperatures from 277 to 293 K in the time interval from 60 to $6·10^3$ s are shown in the form of log–log and normal plots in Fig. 4 and 5, respectively. It is seen in these figures that the experimental data can be described within the error of the measurements by the power law θ ~ $t^{-α}$ with the exponent α depending on the temperature.

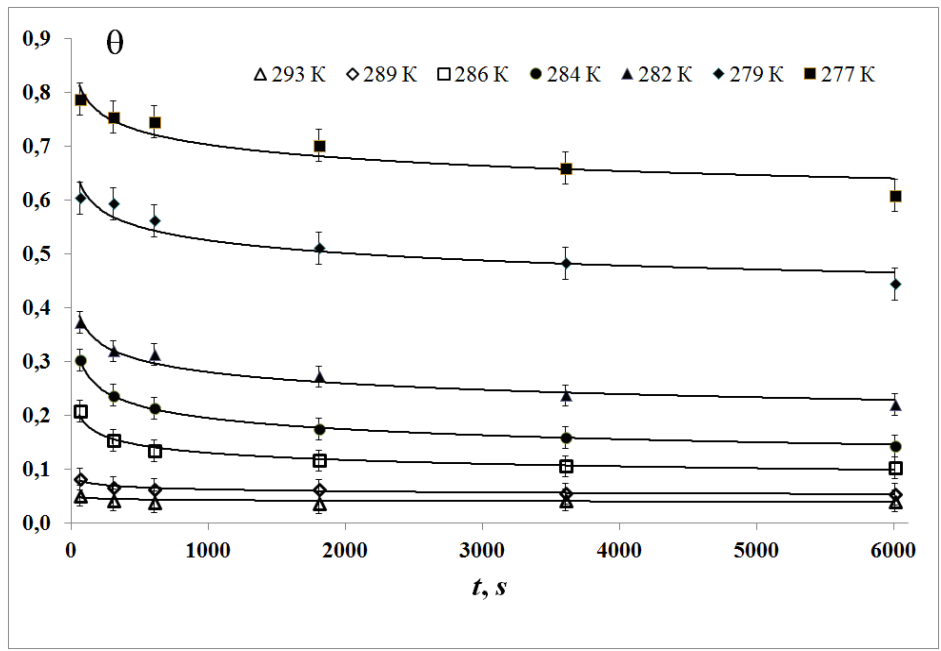

Fig. 4. Dependences θ(t) at seven temperatures from 277 to 293 K in the time interval from 60 to $6·10^3$ s. Experimental points in comparison with the power-law approximations θ ~ $t^{-α}$ shown by solid lines.

It is seen in Fig. 4 that the slope of the linear dependence in the log–log plot increases with an increase in the temperature from 277 to 286 K and, then, decreases with an increase in the temperature to 293 K. Consequently, the exponent α has an extremum at the temperature $T \approx 285$ K, which corresponds to the maximum extrusion rate of the confined liquid at this temperature. The temperature dependence of the exponent α is shown in Fig. 6. The confidence parameter $R^2$ (see Table 1) for the power-law approximation $\theta \sim t^{-\alpha}$ of the experimental values $\theta(t)$ presented in Figs. 4 and 5 was used to estimate the error in the α values.

The small confidence parameter $R^2 = 0.315$ of the approximation at $T = 293$ K can be due to a change in the relaxation mechanism. Relaxation at $\theta < \theta_c$ occurs via the known vapor transport mechanism [42]. In this case, since the hydrodynamic flow time in a channel is independent of the radius of the channel for a Knudsen gas, the relaxation law should be logarithmic rather than power, because the Knudsen number in the experiments at $p = 10^5$ Pa, $\bar{R} = 4\acute{\imath}\grave{\imath}$, and $T \approx 300$ K is $Kn > 10$.

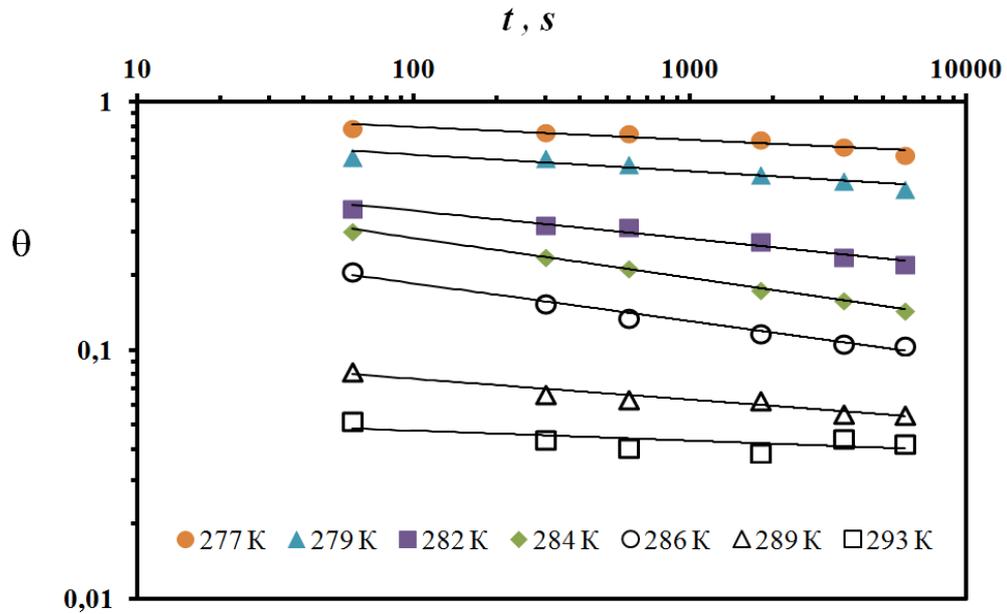

Fig. 5. Log–log plots of θ(t) for seven temperatures from 277 to 293 K in the time interval from 60 to $6 \cdot 10^3$ s. Experimental points in comparison with the power-law approximations $\theta \sim t^{-\alpha}$ shown by solid lines.

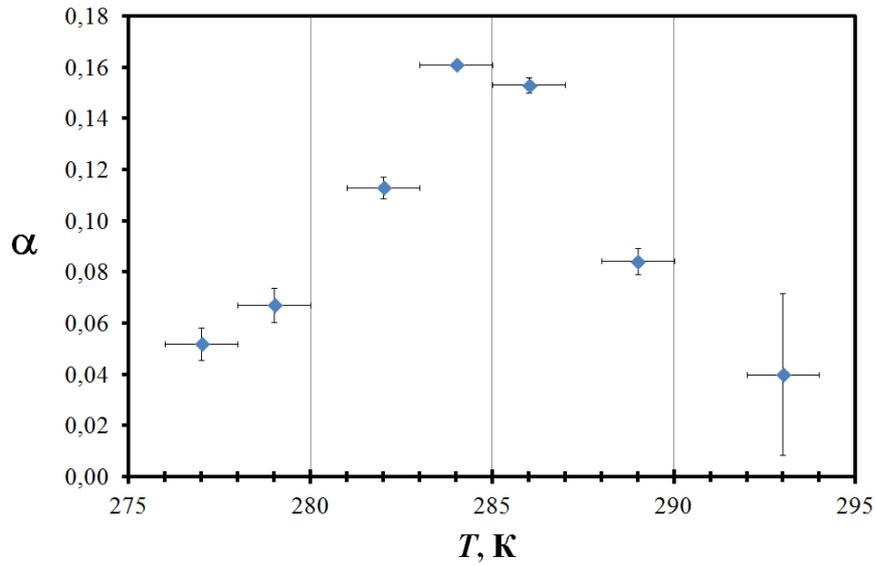

Fig. 6. Temperature dependence of the exponent α. The error in the $y$ axis is defined as $\alpha(T)\cdot(1 - R^2(T))$.

Table 1. Confidence of the power-law approximation $\theta \sim t^{-\alpha}$ of the experimental values $\theta(t)$ presented in Figs. 4 and 5

| $T$, K | $R^2$ |
|---|---|
| 277 | 0.8776 |
| 279 | 0.9003 |
| 282 | 0.9625 |
| 284 | 0.9985 |
| 286 | 0.9814 |
| 289 | 0.9426 |
| 293 | 0.3150 |

The performed experiments provide the following conclusions. When the excess pressure was varied from $450\cdot10^5$ to ~$120\cdot10^5$ Pa, the extrusion of the liquid from the filled porous medium was not observed in the time interval $t = 5$ s in the entire temperature range under study. Extrusion begins at the excess pressure $p < 120\cdot10^5$ Pa. In this case, the extrusion of the confined liquid occurs differently at temperatures above and below the dispersion transition temperature $T = T_d \approx 284$ K. At temperatures $T > 289$ K, the degree of filling changes to values below the percolation threshold $\theta_c$ in filled pores in the time $\tau_p = 10\tilde{n}\hat{a}\hat{e}$ of the reduction of the pressure from $120\cdot10^5$ Pa to zero. In this case, in the time interval from 10 to $10^6$ s, the degree of filling $\theta$ of pores with the liquid varies from 0.08 to 0.02 according to the power law $\theta \sim t^{-a}$ with the exponent $a \leq 0.05$.

At $T < T_d$, the partial extrusion of ~20% of the liquid is observed in the time $\tau_p = 10$ ñ å ê of the reduction of the pressure from $120 \cdot 10^5$ Pa to zero. The degree of filling for the confined liquid $\theta$ is larger than the percolation threshold $\theta_c$, and the fast (in the time $\sim \tau_0 \sim 0.1$ ñ å ê [44]) hydrodynamic extrusion of the liquid through the percolation cluster of filled pores is possble. However, this process does not occur, and the degree of filling in the time from 10 to $6 \cdot 10^3$ s decreases from 0.8 to 0.6 according to the power law $\theta \sim t^{-a}$ with the exponent $a \leq 0.2$.

## 3. STAGES OF RELAXATION

The performed experiments show that two different relaxation scenarios are possible after the filling of the porous medium and the subsequent removal of excess pressure. At the temperatures $T > 284$ K, the most part of the liquid, $\geq 80\%$, is in an unstable state and flows from the porous medium in a characteristic time of the removal of excess pressure $t \sim 10$ s. The remaining part of the liquid with the fraction of filled pores below the percolation threshold $\theta < \theta_c$ at times $t >> 10$ s slowly relaxes according to a power law $\theta \sim t^{-a}$ with a small exponent $a \leq 0.05$. At temperatures $T < T_d$, a fraction of the liquid after filling at excess pressure and the removal excess pressure can be in an unstable state and flows out in the time of the removal of excess pressure $t = 10$ s through filled pores of the infinite percolation cluster. Partial extrusion up to 60% of the liquid occurs in this time. According to [62], the remaining large (~80%) part of the liquid in observation times from 20 to $6 \cdot 10^3$ s is in a metastable state and slowly flows out. In this case, the degree of filling decreases from 0.8 to 0.6 according to a power law $\theta \sim t^{-a}$ with the exponent $a \leq 0.2$ depending on the temperature.

Thus, it is necessary to describe the process of extrusion of the confined liquid at times of reduction of excess pressure $t < 10$ s in two different cases: (i) where the decay of unstable states of filled pores occurs to the degrees of filling $\theta$ below the percolation threshold $\theta_c$ when isolated clusters of pores with the confined liquid can be formed and (ii) where the decay of unstable states ends in the time $t \sim 10$ s at the degree of filling $\theta$ above the percolation threshold $\theta_c$ and results in the formation of a metastable state.

The main aims are to describe the kinetics of relaxation of the metastable state formed in the time $\tau_p = 10$ ñ å ê at $T < T_d$ and $\theta > \theta_c$ and kinetics of its relaxation at observation times $t >> 10$ s and to reveal physical reasons for the anomalously slow, power-law, relaxation of this state.

### 3.1. Kinetics of the Formation of the Metastable State

We consider the dynamics of the extrusion of the nonwetting liquid from the nanoporous medium. The porous medium at the initial time is completely filled at excess pressure $p_0 = 450 \cdot 10^5$ Pa. With a decrease in the pressure down to the critical value $p_{\bar{n}} \sim 120 \cdot 10^5$ Pa, the process of extrusion of the liquid from the porous medium begins. The aim is to calculate the time dependence of the volume fraction of the liquid remaining in the porous medium $\theta(t)$ with the reduction of excess pressure $p(t)$ to zero in the pressure reduction time $\tau_p$.

We assume that the disordered nanoporous medium contains pores with different sizes and the porous medium is much larger than the maximum pore, and the infinite percolation cluster of pores is formed in it so that the porous medium can be considered as infinite.

The extrusion of the liquid from the nanoporous medium can occur through two different mechanisms. If the degree of filling with the confined liquid θ is above the percolation threshold θ_c for filled pores, these pores form a connected system of filled pores through which the liquid can be extruded with the characteristic hydrodynamic time $\tau_0 \sim 0.1 \tilde{n}\tilde{a}\hat{e}$ [44]. If the degree of filling with the confined liquid is θ < θ_c, any connected path for the extrusion of the liquid is absent and the extrusion of the liquid is possible through the mechanism of capillary evaporation–condensation at the boundary of a granule of the porous medium.

The characteristic pressure reduction time in the performed experiments was $\tau_p = 10 \tilde{n}\tilde{a}\hat{e}$; consequently, $\tau_p \gg \tau_0$. The process of extrusion of the liquid at pressure variation times $\tau_p \gg \tau_0$ can be considered as a process occurring in a quasistationary medium with excess pressure $p(t)$ decreasing slowly to zero when $p = (1.00 \pm 0.05) \cdot 10^5$ Pa.

Under these conditions, for the extrusion of the nonwetting liquid from the porous medium filled with the nonwetting liquid at pressure $p$, it is necessary to do work for the depletion of pores of the porous medium. A filled pore in the porous medium can be in states corresponding to different extrusion times depending on its radius [62]. The time of extrusion of the liquid from the pore can be calculated within the statistical theory of fluctuations. We consider a change in the state of the (liquid–porous medium) system at the extrusion of the liquid from the pore surrounded by empty and filled pores at partial filling of the porous medium. These surrounding pores are connected to the pore under consideration by throats in whose mouths a meniscus is formed if one of the two connected pores is not filled with the liquid. According to [66], the probability $w$ of changing the state of the system in unit time at the extrusion of the liquid from the pore under consideration under the action of fluctuations in the system is given by the expression

$$w \sim exp(\Delta S),$$

where $\Delta S$ is the change in the entropy of the system at the extrusion of the liquid from the pore. The proportionality coefficient in this relation is determined by the dynamics of extrusion of the liquid. We assume that a change in the temperature of the system in this process can be neglected. This assumption corresponds to a small thermal effect in the experiments [39, 67]. In this case, the probability can be represented in the form

$$w = w_0 exp(-\delta A / T). \qquad (1)$$

Here, $w_0$ is the pre-exponential factor presenting the dynamics of extrusion of the liquid from the porous medium, $\delta A$ is the isothermal work that should be spent on the extrusion of the liquid from the pore. According to Eq. (1), the time of extrusion of the liquid from the pore is given by the expression

$$\tau = \tau_0 exp(\delta A / T). \quad (2)$$

Here, $\tau_0(R)$ is the hydrodynamic time of extrusion of the liquid determined by the dynamics of extrusion of the liquid from the porous medium. This time can be estimated as follows. If the liquid is extruded from the pore with the radius $R$ through the channel of filled pores with the same radius, $\tau_0 = \dfrac{4\pi R^3}{3Q(R)}$ [62]. In the case of the flow of the liquid through the channel of the radius $R$, $Q(R) \sim R^4$ [68] and

$$\tau_0 \sim 1 / R \ . \quad (3)$$

According to Eq. (2), $\delta A$ serves as the potential barrier for the extrusion of the liquid from the pore. If $\delta A < 0$, the characteristic extrusion time $\tau$ is determined by the hydrodynamic time $\tau_0$ of the motion of the liquid in the porous medium. If $\delta A > 0$, the characteristic liquid extrusion time is determined by the extrusion potential barrier $\delta A$. The characteristic liquid extrusion time in this case can be much larger than the hydrodynamic time $\tau >> \tau_0$ when the barrier is $\delta A > T$. The barrier $\delta A$ includes the work ($pV$) done by the system on an increase in its volume by the volume $V$ of the pore at a pressure $p$ and a change in the surface energy $\delta\varepsilon$ of the liquid in the pore at the extrusion of the liquid. This change in the energy $\delta\varepsilon$ should include, first, a change in the energy $\delta\varepsilon_1$ at the extrusion of the liquid from the pore associated with the interaction of the liquid with the frame of the porous medium and, second, a change in the energy of the environment of the pore $\delta\varepsilon_{int}$ from which the liquid flows. This is because the extrusion of the liquid from the filled pore results in the formation of menisci in the mouths of throats connecting this pore to neighboring liquid-filled pores and in the disappearance of menisci in the mouths of throats connecting this pore to empty pores. As a result, the energy of the liquid in neighboring pores changes in the first and next coordination spheres. This change in the energy depends on the numbers of filled and empty pores in the environment of the pore from which the liquid flows (local configuration) and, therefore, on the degree of filling of the porous medium $\theta$. This can be treated as the <<multiparticle interaction>> of the liquid in the pore with its environment, which depends on the degree of filling $\theta$.

The expression for $\delta A$ in the case of a spherical pore with the radius $R$ can be represented in the form [45]

$$\delta A(R,\theta) = pV + \delta\varepsilon$$

$$\delta\varepsilon = \delta\varepsilon_1(R) + \delta\varepsilon_{int}(R,\theta)$$

$$\delta\varepsilon_1(R) = -\delta\sigma(1-\eta(R))S, \quad \eta(R) = \frac{<S_m(R)>}{S}, \qquad (4)$$

$$\delta\varepsilon_{int}(R,\theta) = \sigma W(z,\theta)\eta(R)$$

Here, $V = \frac{4}{3}\pi R^3$ and $S = 4\pi R^2$ are the volume and area of the pore, respectively; $<S_m>$ is the mean area of menisci in the local configuration; $\delta\sigma$ is the change in the specific energy of the surface of the solid (frame of the porous medium) at the extrusion of the liquid; and $\sigma$ is the specific energy of the (nonwettingая liquid–gas) interface.

It follows from Eqs. (4) and (5) that $\delta\varepsilon_1$ depends on a change in the surface energy $\delta\sigma$ at the extrusion of the liquid from the pore and, therefore, takes into account the interaction of the liquid with the frame of the porous medium. The quantity $\varepsilon_{int}$ is determined as a change in the surface energy of menisci in the mouths of pores in the environment of the pore from which the liquid flows. For this reason, $\delta\varepsilon_{int}(R,\theta_1)$ can be treated as the effective «multiparticle interaction» of the liquid cluster in the pore with neighboring pores.

The function $\eta(R)$ defined as the ratio of the mean area of menisci to the area of the surface of the pore was calculated in [69] for the medium having pores with different radii with allowance of correlations in the spatial arrangement of pores in the medium, $\eta \sim q(R_0/R)^{-\alpha}, \alpha \approx 0.3, q \sim 1$. The quantity $R_0$ is the minimum radius of pores in the porous medium, which is determined by the pore radius distribution function and is in order of magnitude the mean radius of pores divided by the mean number of the nearest neighbors, $R_0 \sim \overline{R}/\overline{z}$.

The function $W(z,\theta_1)$ defined as the average difference between the number of menisci after and before the depletion of the pore is given by the expression

$$W(z,\theta) = \sum_{n=0}^{z-1}(1-\theta)^n(P(\theta))^{z-n}\frac{z-2n}{z}\frac{z!}{n!(z-n)!} \cdot (5)$$

Here, $P(\theta)$ is the probability that the filled pore belongs to the infinite cluster of filled pores. The first and second factors correspond to the probability that an empty pore is near the infinite cluster of filled pores under the condition that this pore is surrounded by $n$ empty and $z-n$ filled pores and, consequently, contains $n$ menisci. The third factor determines the difference between the relative numbers of menisci after $(z-n)$ and before $(n)$ the filling of the pore. The combinatory factor presents the variants of the distribution of $n$ menisci over the nearest neighbors of the given pore and corresponds to the degeneracy of the local geometric state of configurations of the pore and filled pores in the first coordination sphere. Thus, each term in sum (5) describes a change in the number of menisci at a fixed relation between the numbers of filled and empty pores. Summation in Eq. (5) includes all possible variants of the mutual arrangement of empty and filled pores and makes it possible on average to take into account variations of local configurations of the pore and its environment consisting of filled and empty pores in the disordered porous medium. The number of the nearest neighbors $z$ is related to the porosoity $\varphi$ and can be calculated for various models of the porous medium [70]. In particular, within the model of randomly distributed spheres, $z = -8\ln(1-\varphi)$ [70]. According to Eq. (5), the function $W(z, \theta_1)$ depends slightly on a particular choice of the probability $P(\theta)$ that a filled pore belongs to the infinite cluster of filled pores [62].

According to Eqs. (4) and (5), the energy of the «multiparticle interaction» of the liquid cluster with clusters in neighboring connected pores, together with the surface energy of the interaction of liquid clusters with the frame of the medium and the work $pV$, forms the local random profile of a potential barrier for the extrusion of the liquid from the pore in the disordered medium. In view of the dependence of the number of the filled (empty) pores on the degree of filling, this barrier can be negative for some pores and the liquid rapidly flows from them [25, 62]. The barrier is positive for the other pores and metastable states appear. It follows from Eqs. (4) and (5) that the energy of the «multiparticle interaction» is positive, $\delta\varepsilon_{int} > 0$, at $\theta_0 \leq \theta \leq 1$ and is negative, $\delta\varepsilon_{int} < 0$, at degrees of filling of the porous medium below a certain value $\theta_c < \theta < \theta_0$ [25, 62]. According to estimates [24, 25], $\theta_0 \approx 0.3$ for the porous medium with a narrow pore size distribution with the relative width $\Delta R / \bar{R} << 1$.

It follows from Eq. (4) that the potential barrier $\delta A$ is determined by the pressure $p$ and by the surface energy $\delta\varepsilon$ of the liquid in the pore at the extrusion of the liquid and depends on the radius of the pore $R$. According to Eq. (4), the quantity $pV$ is always positive, whereas the surface energy $\delta\varepsilon$ has a maximum $\varepsilon_{max}$ at the radius of the pore $R = R_{max}(z, \theta_1)$ and can change sign at radii $R*(z, \theta)$ for which $\delta\varepsilon_{int}(R, \theta_1) = | \delta\varepsilon_1 |$. It follows from this condition that

$$R^*(z, \theta_1) = q^{\frac{1}{\alpha}} R_0 (1 + \frac{\sigma}{\delta\sigma} W(z, \theta_1))^{\frac{1}{\alpha}}, \qquad (6)$$

where $R_0 \sim \bar{R} / \bar{z}$ is the minimum size of pores in the pore volume distribution $f^V(R)$. According to Eqs. (4) and (6), the surface energy at the radius of the pore $R < R^*$ $(z, \theta)$ is positive, $\delta\varepsilon > 0$. In this case, the potential barrier $\delta A$ is positive at any pressure $p$. Consequently, the extrusion time from such a state given by Eq. (2) is exponentially large, $\tau = \tau_0 exp(\delta A / T) >> \tau_0$, compared to the hydrodynamic time $\tau_0 \sim 0.1 n\tilde{a}\hat{e}$.

The surface energy $\delta\varepsilon$ at $R > R^*$ $(z,\theta)$ is negative, $\delta\varepsilon < 0$. It follows from Eq. (4) that the barrier $\delta A$ in this case critically depends on the pressure $p$. The barrier $\delta A$ is positive, $\delta A > 0$, at pressures $p > p_c = \dfrac{\delta\varepsilon}{V}$ and is negative, $\delta A < 0$, at $p < p_c = \dfrac{\delta\varepsilon}{V}$. The critical pressure depends on the radius of the pore $p_c(R)$. For this reason, at excess pressure $p$, the barrier for some pores is negative and, according to Eq. (2), the liquid can flow from these pores in the hydrodynamic time $\tau_0 \sim 0.1\,нс$. The barrier for the other pores is positive, $\delta A > 0$. According to Eq. (2), the extrusion time from such pores can be exponentially large, $\tau = \tau_0 exp(\delta A / T) >> \tau_0$.

It follows from Eq. (6) that the quantities $R^*(z,\theta_1)$ and $\varepsilon_{max}(z,\theta_1)$ depend on the temperature because of the temperature dependence of the surface tension coefficients $\sigma(T)$ and $\delta\sigma(T)$. The analysis shows that the quantities $R^*(z,\theta_1)$ and $\varepsilon_{max}(z,\theta_1)$ increase with a decrease in the temperature and with an increase in the degree of filling $\theta_1$ [25]. Correspondingly, the potential barrier $\delta A$ also depends on the temperature, increasing with a decrease in the temperature and with an increase in the degree of filling $\theta$.

Figure 7 shows the dependence of the potential barrier $\delta A$ on the radius of the pore at the complete filling $\theta = 1$ at $T = 279$ K and at different pressures as calculated from Eqs. (4) and (5) with the use of the results [71] for the surface tension coefficient $\sigma(T)$ and its temperature dependence. The surface tension coefficient of water at $T = 293$ K is 75 mJ/m$^2$ [71]. The quantity $\delta\sigma$ and its temperature dependence for the system under study were determined from the temperature dependence of the extrusion pressure by the method described in [57]. The $\delta\sigma$ value at $T = 293$ K is 22 mJ/m$^2$. The quantity $R_0 \sim \dfrac{\bar{R}}{\bar{z}}$ was estimated within the model of randomly distributed spheres [24, 25]. At the porosity $\varphi \sim 0.5$ and $\bar{R} \sim 5\,нм$, $R_0 \sim 1\,нм$.

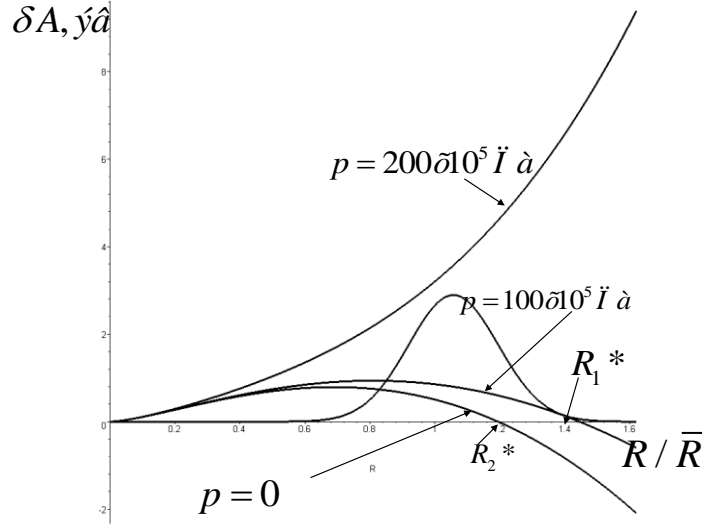

Fig. 7. Potential barrier $\delta A$ versus the radius of the pore for complete filling $\theta_1 = 1$ at $T = 277$ K and different pressures; $\overline{R} = 5$ í ì and the pore volume distribution function $f^V(R)$ with $\dfrac{\Delta R}{R} = 0.1$ and $z = 6$, $R_1*$ and $R_2*$ are radii above which $\delta A < 0$ for the excess pressure $p = 100 \cdot 10^5$ Pa and $p = 0$, respectively.

It is seen in Fig. 7 that the energy barrier for extrusion from pores with the distribution function $f^V(R)$ at pressures $p > 200 \cdot 10^5$ Pa is $\delta A \sim 4$ ýâ. The reduction of the pressure to $p \sim 100 \cdot 10^5$ Pa is accompanied by a change in the potential barrier. The potential barrier for extrusion from pores with the radius $R > R_1*$ disappears, $\delta A(R > R_1*) \leq 0$ (Fig. 7). The potential barrier $\delta A(R < R_1*)$ for extrusion from pores with the radius $R < R_1*$ is positive and increases within the pore volume distribution $f^V(R)$.

The dependence of the volume fraction of the liquid remaining in the porous medium at the reduction of the pressure at times $t <\sim \tau_p \sim 10$ ñåê can be calculated taking into account that extrusion at times $\tau_p \gg \tau_0$ can be considered as a quasistatic process at excess external pressure $p(t)$ decreasing slowly to zero. In this case, following [44], the probability of finding the pore in the filled state at the pressure $p(t)$ can be determined as

$$w_o\big(p(t), R\big) = \left[1 + \exp\left(\dfrac{-\delta A\big(p(t), R\big)}{T}\right)\right]^{-1}. \qquad (7)$$

Here, the quantity $\delta A$ serving as the potential barrier for extrusion is defined by Eq. (4). In this case, the time dependence of the volume fraction of the liquid remaining in the porous medium upon reduction of the pressure at times $t < \tau_p = 10$ ñåê can be represented in the form

$$\theta(t) = \int\limits_0^\infty w_o(p(t), R) f^V(R) dR , \quad (8)$$

where $f^V(R)$ is the pore volume distribution function.

According to Eq. (8), the time dependence of the volume fraction of the remaining liquid at times $t < \tau_p = 10 \tilde{n}\hat{a}\hat{e}$ is determined by Eqs. (4) and (7). The potential barrier at nonzero excess pressure $p(t) \neq 0$ is determined by the competition between a change in the surface energy $\delta\varepsilon$ of the liquid in the pore and the work necessary for the depletion of the pore with the volume $V$ at excess pressure $p(t)$. At $t = \tau_p = 10 \tilde{n}\hat{a}\hat{e}$, excess pressure $p(t)$ vanishes and the potential barrier for the extrusion of the liquid is determined by the competition between the energy of the interaction of the liquid with the frame of the porous medium $\delta\varepsilon_1$ and the energy of the effective «multiparticle interaction» of the liquid cluster in the pore with liquid clusters in neighboring pores $\delta\varepsilon_{int}$. According to Eq. (4), liquid clusters for which the energy of the effective «multiparticle interaction» with liquid clusters in neighboring pores is positive, $\delta\varepsilon_{int} > 0$, and $\delta\varepsilon_{int} > | \delta\varepsilon_1 |$ form the metastable state of the system corresponding to the bound states of interacting local configurations of filled pores [62].

### 3.2. *Kinetics of the Relaxation and Decay of the Metastable State at* $\theta > \theta_c$

We now determine the character of relaxation of this state $\theta(t)$ at times $t > \tau_p$ in the absence of excess pressure $p = 0$ and $\theta(t) > \theta_c$. In this case, the system includes the percolation cluster of filled pores through which the liquid flows from the porous medium. According to Eq. (5), the effective «multiparticle interaction» of the liquid cluster with neighboring clusters in pores is nonzero and is attractive at $\theta > \theta_0 \sim 0.2$. This interaction ensures the existence of the metastable state of some pores satisfying the condition $\delta\varepsilon_{int} > | \delta\varepsilon_1 |$. It follows from Eqs. (2)–(4) that the time of extrusion of the liquid from the pore in the metastable state of the system is

$$\tau = \tau_0 exp(-\delta\varepsilon(R, \theta, T)/T), \tau_0 \sim 1/R . \quad (9)$$

This time is determined by the degree of filling $\theta$ of the porous medium with the liquid in the metastable state, the radius of the pore $R$, and the temperature $T$.

Following [62], we introduce the distribution function $F(t)$ in the time of extrusion from the pore with the radius $R$ and, correspondingly, the volume $V = (4\pi/3)R^3$. This function determines the volume fraction of pores $d\theta(t)$ from which the liquid flows in the time $dt$: $d\theta(t) = F(t)dt$. For the nonrandom liquid extrusion time $\tau$, the function is $F^V(t) = \delta(t - \tau)$, where $\delta(t)$ is the Dirac delta function. Since the porous medium contains pores with different volumes with the probability determined by the pore volume distribution function, it follows from Eq. (9) that the height of the barrier $\varepsilon(R, \theta_1)$, pre-exponential factor $\tau_0(R)$, and extrusion time $\tau(R)$ are random functions. In this case, the distribution function in the time of extrusion of the liquid from the porous medium $F^V(t)$ has the form

$$F^V(t) = \int_0^\infty \delta(t - \tau\,(R)) f^V(R) dR\,.\qquad (10)$$

Here, $\tau\,(R)$ is given by Eq. (9) and $f^V(R)$ is the volume distribution function of pores normalized to unity. It follows from Eqs. (4), (9), and (10) that the time of flow of the liquid is not the same for all pores and the number of liquid clusters involved in extrusion depends on the time. As will be shown below, this can result in the power-law character of relaxation of the metastable state.

According to Eq. (10), the relaxation function $F^V(t)$ of this state $\theta(t)$ is normalized to unity, $\int_0^\infty d\tau F^V(\tau) = 1$, and the integral

$$\int_0^t d\tau F^V(\tau) = \theta(t) \qquad (11)$$

determines the volume fraction of pores $\theta(t)$ from which the liquid flows in the time interval from zero to $t$. Since we describe the relaxation of the metastable state formed at the time $t = \tau_p$, we measure the relaxation time from this time. Then, if the medium at the initial time (time of the formation of the metastable state) is filled to the degree of filling $\theta(0) = \theta$, the volume fraction of pores $\theta(t)$ from which the liquid did not flow to the time $t$ is determined by the expression

$$\theta(t) = \theta \int_t^\infty dt F^V(t)\,.\qquad (12)$$

Calculating integral (12), we find that the distribution function $F^V(t)$ has the form

$$F^V(t) = f^V(R(t)) \frac{dR(t)}{dt}\,.\qquad (13)$$

Here, $R(t)$ is the solution of the equation

$$\tau\,(R(t)) = t\,.\qquad (14)$$

Integral (12) of the distribution function given by Eq. (13) determines the time dependence of the volume fraction of the liquid in the porous medium in the metastable state. To solve Eq. (14), we note that the metastable state is formed from pores with the volume distribution function $f^V(R)$ and radius $R < R^*$, where $R^*$ is given by Eq. (6). In the case of pore distributions $f^V(R)$ with the relative width $\Delta R / \bar{R} < 0.5$, ~90% of the volume of the liquid is in the pores with the radii $R < \bar{R} + \Delta R$. Therefore, ~10% of the liquid is in the pores with the radii $R > \bar{R} + \Delta R$. Liquid clusters in pores with the radii $R > R^*$ are not in a metastable state: the liquid from such pores flows in a hydrodynamic time of ~ $\tau_0$.

For this reason, we assume for simplicity that the distribution function of pores with liquid clusters in the metastable state is constant at $0 < R < R*$ and vanishes at $R > R*$. Then, in view of the normalization, the volume distribution function of pores in this approximation $\tilde{f}^V(R)$ has the form

$$\tilde{f}^V(R) = \begin{Bmatrix} \dfrac{1}{R*}\displaystyle\int_0^{R*} f^V(R)dR, 0 < R < R* \\ 0, R > R* \end{Bmatrix} \tag{15}$$

The analysis indicates that the barrier $\varepsilon(R,\theta)$ for pores in this radius interval is a smooth function of the radius. In this case, in the radius interval $0 < R < R*$ of the pores constituting the metastable state, the characteristic extrusion time $\tau(R)$, according to Eqs. (13) and (14), can be represented in the form

$$\tau(R) = \tau_0 (\frac{\overline{R}}{R}) \exp(\delta\varepsilon / T) = \tau_0 (\frac{\overline{R}}{R}) \exp(\varepsilon_{\max} / T) \exp((\delta\varepsilon - \varepsilon_{\max}) / T) \approx$$

$$\approx \tau_0 \exp(\varepsilon_{\max} / T) \exp(-\delta\varepsilon''(R - R_{\max})^2 / 2T - \ln(\frac{R}{\overline{R}})) \approx \tau_0 \exp(\varepsilon_{\max} / T) \exp(-\ln(\frac{R}{\overline{R}})(\frac{\delta\varepsilon''(\Delta R)^2}{2T\ln(1+\frac{\Delta R}{\overline{R}})} + 1)) =$$

$$= (\frac{\overline{R}}{R})^\beta \tau_0 \exp(\varepsilon_{\max} / T),$$

$$\beta = 1 + \frac{\delta\varepsilon''(\Delta R)^2}{2T\ln(1+\frac{\Delta R}{\overline{R}})},$$

$$\delta\varepsilon'' = (\frac{d^2\delta\varepsilon(R)}{dR^2})_{R=R_{\max}}$$

$$\tag{16}$$

Here, $\varepsilon_{\max}(\theta_1)$ is the maximum height of the barrier $\varepsilon(R,\theta_1)$ (4) and $\tau_0$ is the hydrodynamic time of extrusion of the liquid from pores of the porous medium [62]. The exponent $\beta$ for $\frac{\Delta R}{\overline{R}} << 1$ follows from Eqs. (4) and (16) in the form

$$\beta = 1 + (2-\alpha)\frac{\Delta R}{\overline{R}}\frac{\varepsilon_{\max}}{T}. \tag{17}$$

Here, $\alpha$ is the exponent in the coefficient of connection $\eta \sim q(R_0 / R)^{-\alpha}, \alpha \approx 0.3$.

Using the volume distribution function given by Eq. (15) and Eqs. (12) and (13), we obtain $\theta(t)$ at the times $t > \tau_p$ in the form

$$\theta(t) \sim \theta_p (\frac{\tau_{q1}}{t})^a, a = \frac{1}{1 + (2-\alpha)\frac{\Delta R}{\overline{R}}\frac{\varepsilon_{\max}}{T}}$$

$$\tau_{q1} \sim \tau_0 \exp(\frac{\varepsilon_{\max}}{T}), \theta_p = \frac{\overline{R}}{R*}\int_0^{R^*} dR f^V(R)$$

$$\tag{18}$$

It follows from Eq. (18) that the volume fraction of the remaining liquid in the metastable state at times $t > \tau_p$ decreases according to a power law with the exponent $a$ and characteristic time $\tau_{q1} \sim \tau_0 \exp(\frac{\varepsilon_{max}}{T})$.

According to Eqs. (4) and (6), the quantities $R^*(z, \theta_1)$ and $\varepsilon_{max}(z, \theta_1)$ decrease with a decrease in the degree of filling $\theta$ because of the reduction of the energy of the «multiparticle attraction» between local configurations [25]. For this reason, it follows from Eq. (18) that the regime of relaxation changes with a decrease in the degree of filling $\theta(t)$ according to Eq. (18). The relaxation of the metastable state according to Eq. (18) with the exponent $a \ll 1$ for degrees of filling $\theta \sim 1$ at low temperatures and $\frac{\Delta R}{\bar{R}} \frac{\varepsilon_{max}}{T} \gg 1$ is replaced at $\frac{\Delta R}{\bar{R}} \frac{\varepsilon_{max}}{T} < 1$ by fast relaxation with the exponent $a \sim 1$ and characteristic relaxation time $\tau_{q1} \sim \tau_0 \exp(\frac{\varepsilon_{max}}{T}) \sim \tau_0 \exp(\frac{\bar{R}}{\Delta R}) < 1000\tau_0 \sim 100\,\text{s}$. This corresponds to the decay of the formed metastable state. According to Eq. (18), the decay of the metastable state begins at times for which the degree of filling $\theta(t)$ is such that $\frac{\Delta R}{\bar{R}} \frac{\varepsilon_{max}}{T} \sim 1$.

Expression (18) indicates that the power-law dependence $\theta(t)$ for the relaxation and decay of the metastable state is due to the polydisperse distribution ($\Delta R / \bar{R}$) of pores in the porous medium and to the interaction between local configurations whose maximum energy $\varepsilon_{max}$ determines both the characteristic relaxation and decay time $\tau_{q1}$ and the exponent $a$.

### 3.3. *Discussion of the Results*

We now discuss the time dependences от the volume of the liquid remaining in the porous medium at three stages of the formation, relaxation, and decay of the metastable state.

It follows from Eq. (8) that the time dependence of the volume fraction of the remaining liquid at the times $t < \tau_p = 10$ ñåê is determined by Eqs. (2), (4), and (7). Relations (4) and (5) give the potential barrier for extrusion of the liquid from the pore with the radius $R$ in the porous medium filled with the nonwetting liquid at the pressure $p$ (Fig. 1). At the pressure $p \sim 200 \cdot 10^5$ Pa and the radius of the pore $R \sim 3$ íì , $pV \sim 3.5$ ýâ.

Estimates show that the maximum value $\varepsilon_{max}$ for the system under study at the temperatures below the transition dispersion temperature $T = 277$ K $< T_d = 284$ K is $\varepsilon_{max} \geq 0.8$ ýâ. At excess pressures $p > 200 \cdot 10^5$ Pa, the energy barrier for extrusion from pores with the pore distribution $f^V(R)$ is $\delta A \sim 4\acute{y}\hat{a}$ (Fig. 7). For this reason, liquid clusters in all pores at $p > 200 \cdot 10^5$ Pa are in states with the extrusion barrier $\delta A \sim 4\acute{y}\hat{a}$. In this case, the time of extrusion of the liquid from pores is estimated by Eqs. (2) and (4) as $\tau > 10^{33} \tilde{n}\hat{a}\hat{e}$. When excess pressure is reduced to $p \sim 100 \cdot 10^5$ Pa, the barrier $\delta A$ becomes negative for liquid clusters in pores with the radiii $R > R_1^*$ (Fig. 7). The liquid flows from such pores in the hydrodynamic time $\tau \sim 10^{-1} \tilde{n}\hat{a}\hat{e}$. As the pressure is reduced, the quantity $R^*$ decreases and the number of clusters from which the liquid flows increases. The volume of the extruded liquid at zero excess pressure is determined by the expression

$$\theta = \int_{R^*}^{\infty} f^V(R) dR. \tag{19}$$

This expression makes it possible to calculate the fraction of the remaining liquid at a fixed excess pressure $p < 100 \grave{a}\grave{o}\grave{i}$.

An increase in the temperature results in a decrease in $R^*$. For this reason, according to Eq. (19), as the temperature is increased, the volume of the liquid extruded in the time of vanishing excess pressure increases and, therefore, the volume of the liquid remaining in the porous medium for this time decreases.

According to Eqs. (4) and (8), as excess pressure is reduced from $p > 200 \cdot 10^5$ Pa to about $100 \cdot 10^5$ Pa, the liquid remains in the filled porous medium. A further reduction of the pressure at times $t \le \tau_p = 10 \tilde{n}\hat{a}\hat{e}$ results in the extrusion of the liquid from pores for which $\delta A \le 0$. The volume of the extruded liquid $1 - \theta$ in the excess pressure vanishing time can be estimated from Eq. (19) with $R_2^*$ given by Eq. (6). These estimates by Eqs. (6) and (19) with the dependence $\sigma(T)$ from [71], as well as $\delta\sigma$ and its temperature dependence determined by the method described in [57], give $1 - \theta(T = 277\text{K}) \sim 0.15$, at the temperature $T = 279$ K and $1 - \theta(T = 289\text{K} \sim 0.9$ at the temperature $T = 289$ K. These values are close to the experimental values $1 - \theta = 0.1$ and 0.8 at the time $t = 10 \tilde{n}\hat{a}\hat{e}$, respectively.

It follows from Eqs. (4) and (9) that the character of relaxation of the state formed at the reduction of excess pressure to $p = 0$ is determined by the potential barrier $\delta\varepsilon$. Figure 8 shows the dependence of the potential barrier $\delta\varepsilon$ for the system under consideration at the time $t = \tau_p \sim 10 \tilde{n}\hat{a}\hat{e}$ and the temperatures $T = 277$ and 293 K as calculated by Eq. (4) with the $\sigma, \delta\sigma, \bar{R}, R_0, z, \varphi$ values same as in the case corresponding to Fig. 7.

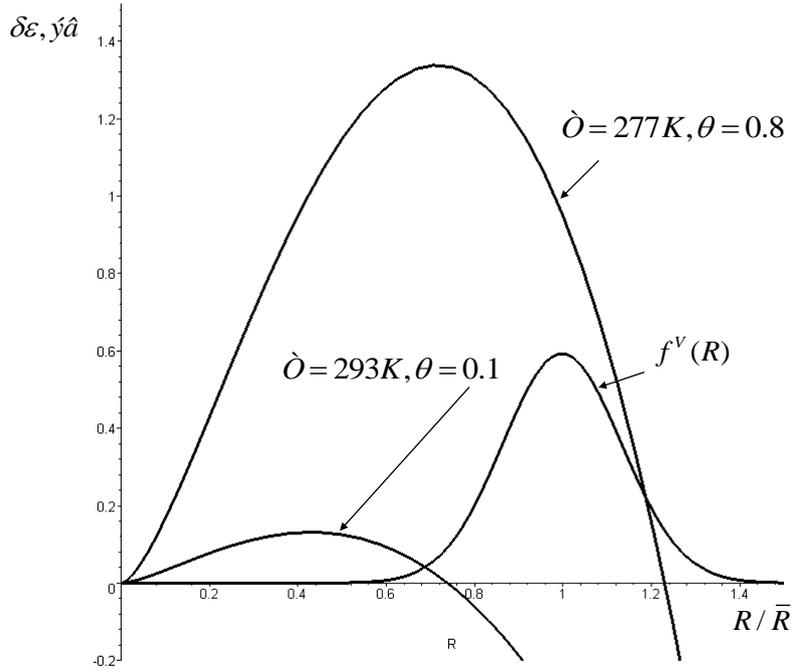

Fig. 8. Potential barrier $\delta\varepsilon$ for the system under consideration at the time $t = \tau_p \sim 10$ сек and temperatures $T = 277$ and 293 K.

It is seen in Fig. 8 that the potential barrier $\delta\varepsilon$ decreases with an increase in the temperature and a decrease in the degree of filling. The potential barrier at $T = 293\text{K}$ and $\theta = 0.1$ (Fig. 4) is below or about the temperature for all pores. In this case, $\theta < \theta_c$ and only isolated clusters of filled pores are formed in the porous medium. Paths for the extrusion of the liquid from these clusters through filled pores are absent. A possible mechanism of the extrusion of the liquid from clusters of filled pores in this case can be recondensation.

At $T = 277$ K and times $t = \tau_p$ when $p = 0$, $\theta > \theta_c$ (Fig. 4) and the potential barrier for the extrusion of the liquid is determined by the competition between the energy of the interaction of the liquid with the frame of the porous medium $\delta\varepsilon_1$ and the energy of the effective «multiparticle interaction» of the liquid cluster in the pore with neighboring pores $\delta\varepsilon_{int}$. For pores with the radii $R < R^*$, the energy of the effective «multiparticle interaction» of the liquid cluster in the pore with clusters in the neighboring pores is $\delta\varepsilon_{int} > 0$ at $\theta \sim 0.8$ larger than $\theta_0 \sim 0.2$ at which the quantity $\delta\varepsilon_{int}$ changes sign and corresponds to repulsion, rather than attraction, between clusters. If $\delta\varepsilon_{int} > \mid \delta\varepsilon_1 \mid$ in this case, the potential barrier for extrusion is positive ( $\delta\varepsilon > 0$ ) and such pores form a metastable state of the system of clusters in pores, which corresponds to bound states of interacting local configurations of filled pores.

According to Eq. (16), the relaxation of the formed metastable state occurs through extrusion from local configurations of pores with radii smaller than $R*$. The energy barrier for extrusion from these configurations $\varepsilon(R, \theta_1)$ is determined by the interaction between local configurations. Such local configurations are «strongly coupled» with their environment and, therefore, «interact» with each other. For this reason, it can be thought that interacting local configurations of the pore and its environment in the case under consideration are condensed and form the metastable state of the entire system of clusters, which decays according to power law (18) with the characteristic time $\tau_{q1} \sim \tau \exp(\frac{\varepsilon_{max}}{T})$, which is $\tau_{q1} \sim 10^5$ ñåê at $\tau \sim 0.1$ ñåê, $\varepsilon_{max} \sim 0.5$ ýâ, and $T = 277 K$. It follows from Eq. (18) that the exponent $a$ for narrow distributions with the relative width $\frac{\Delta R}{R} \sim 0.1$ at these parameters is $a \sim 0.1$.

In view of Eq. (18), the decay of this metastable state is accompanied by a decrease in the degree of filling $\theta$. According to Eq. (4), this results in a decrease in the energy of the interaction between local configurations and, as a result, in a decrease in the energy barrier for extrusion $\varepsilon(R, \theta_1)$. As follows from Eq. (18), the quantity $\varepsilon_{max}$ decreases and the exponent $a$ in Eq. (18) increases. For this reason, the decay rate of this metastable state increases with time. According to Eq. (18), the decay of the metastable state begins when the degree of filling $\theta(t)$ is such that $\frac{\Delta R}{R}\frac{\varepsilon_{max}}{T} \sim 1$. The decay onset time of the metastable state at $\Delta R / \bar{R} \sim 0.1, T = 277 K$ is estimated as $> 10^4$ s in agreement with the experimental data (Fig. 4).

It follows from Eq. (18) that the maximum volume fraction $\theta_q$ of configurations in the metastable state is determined by the integral of the distribution function $\theta_q \sim \int_0^{R*} dR f^V(R)$ and is $\theta_q \sim 0.8$ for a narrow pore volume distribution with the relative width $\frac{\Delta R}{\bar{R}} \sim 0.1$.

The relaxation of the metastable state of interacting local configurations of filled pores continues as long as the percolation cluster of filled pores exists in the system. The existing time of the percolation cluster of filled pores can be estimated from Eq. (18) by equating $\theta(t)$ to the percolation threshold $\theta(t) = \theta_c$. According to Eq. (18), the time of existence of the percolation cluster of filled pores is $t_p \sim \tau_{q1}(\frac{\theta_p}{\theta_c})^{\frac{1}{a}}$. At $a \sim 0.2, \theta_p \sim 0.8, \theta_c \sim 0.1$, with allowance for a change in the degree of filling $\theta$ in the process of decay of the metastable state, this time is $t_p \sim 10^6$ s. At the degree of filling $\theta$ below the percolation threshold, $\theta < \theta_c$, isolated clusters of filled pores are formed in the porous medium. A possible mechanism of extrusion of the liquid from such isolated clusters of filled pores can be recondensation, i.e., capillary evaporation and subsequent capillary condensation at the interface between the porous medium and environment liquid considered in [58, 59]. An analysis similar to that used when deriving Eqs. (16) and (18) gives in this case a logarithmic law of the extrusion of the liquid.

To summarize, the performed analysis shows that there are three stages of the extrusion of the nonwetting liquid from the nanoporous medium.

When the pressure is reduced from $p > 200 \cdot 10^5$ Pa to $p \sim 100 \cdot 10^5$ Pa, liquid clusters in all filled pores in the water–L23 system should be in states with the extrusion barrier $\delta A \sim 4ev$. The time of extrusion from such pores is $\tau > 10^{33} n\mathring{a}\hat{e}$. For this reason, the liquid should remain in the porous medium for the time of pressure reduction to $100 \cdot 10^5$ Pa.

When the pressure is reduced to $p < 100 \cdot 10^5$ Pa, pores appear for which the extrusion barrier $\delta A$ is negative or can be about the temperature $T$. The liquid can be extruded from these pores in the hydrodynamic time $\tau_0 \sim 0.1 n\mathring{a}\hat{e}$ and, therefore, extrusion should be observed for the excess pressure vanishing time $t \sim \tau_p$. As excess pressure vanishes, the number of such pores increases.

Figure 9 shows the qualitative dependences of the volume fraction of the remaining liquid for various temperatures according to Eqs. (7), (8), and (18). It is seen in Fig. 9 that the degree of filling at temperatures $T < T_d = 284$K under the reduction of excess pressure from $100 \cdot 10^5$ Pa to zero in 10 s decreases from $\theta = 1$ to $\theta > \theta_c$ and the loosely bound state decays with the formation of the initial metastable state of the confined liquid. It follows from (8) that the characteristic decay time of the loosely bound state is $\tau_\theta \sim \theta / \dot{\theta} \sim w / \dot{w} \sim \tau_p (T / \delta A)$. Since $\delta A \leq T$ for the loosely bound state, $\tau_\theta$ is larger than the characteristic pressure variation time $\tau_p$, i.e., $\tau_\theta > \tau_p$, and increases with the temperature (Fig. 9a) because of the reduction of the potential barrier $\delta \varepsilon$ at excess pressure $p = 0$ (Fig. 8).

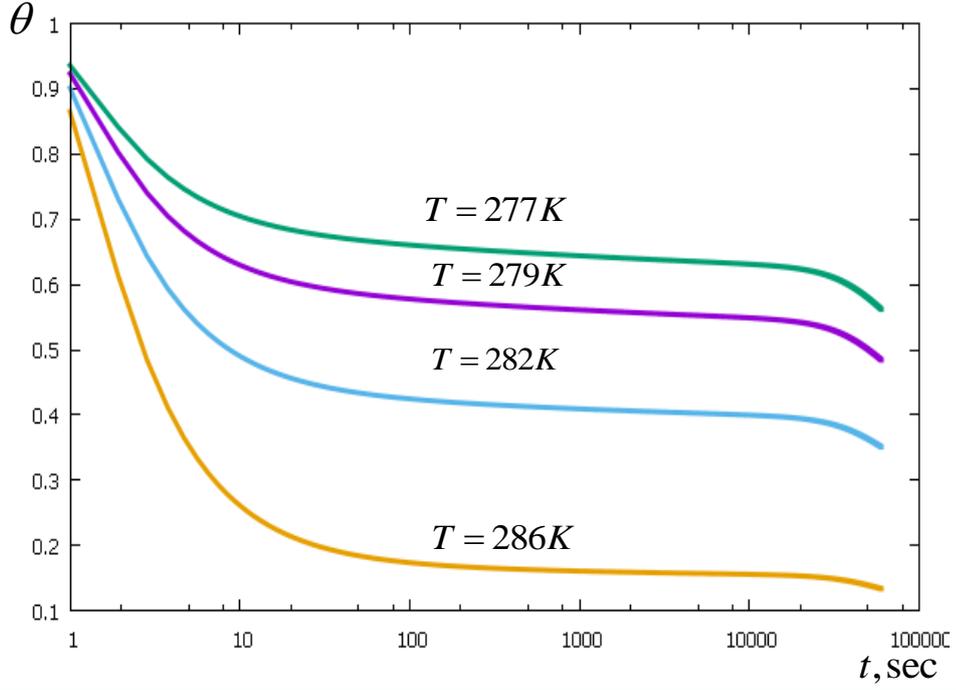

Fig. 9. Qualitative dependence of the volume fraction of the remaining liquid for the temperatures $T < T_d = 284\text{K}$ and $T = 286\text{K}$ according to Eqs. (7), (8), and (18).

It is seen in Fig. 9 that the relaxation of the formed metastable state of the confined liquid occurs in times $t = 100$–$6000$ s according to the power law $\theta(t) \sim t^{-a}$ (18) with the exponent $a \sim 0.1$. In times $t > 10^4$ s, the metastable state decays (see Fig. 9). In this case, the exponent $a$ increases to $a \sim 0.3$. Figure 9 demonstrates that the presented dependence qualitatively describes the experimental data (Fig. 3).

The reduction of excess pressure to zero at the temperatures $T > T_d = 284\text{K}$ results in a decrease in the fraction of the remaining liquid because of the decay of the loosely bound states from $\theta = 1$ to the degrees of filling below the percolation threshold $\theta_c \sim 0.15$. In this case, the relaxation of the system can occur through the evaporation–condensation mechanism, which results in an increase in the characteristic extrusion time of the liquid from the loosely bound states owing both to an increase in the extrusion barrier with the temperature and to an increase in the transfer time of the Knudsen gas from an evaporated pore to the surface of the granule. Figure 9 shows the dependence of the volume fraction of the remaining liquid for the temperature $T = 286\text{K}$ as calculated from Eqs. (7) and (8).

According to Eq. (16), the exponent $a$ depends on the maximum potential barrier and temperature. The exponent $a$ increases with the temperature because of a decrease both in the maximum potential barrier $\varepsilon_{\max}$ and in the ratio $\varepsilon_{\max}/T$. At high temperatures $T > 284$ K, when the excess pressure is reduced to zero, the degree of filling of $\theta$ in the time $t \sim \tau_p$ becomes lower than the percolatoin threshold $\theta_c \sim 0.15$. In this case, the relaxation mechanism of the system changes. The percolation cluster of filled pores disappears and the extrusion of the liquid occurs through a slower evaporation–condensation mechanism. For this reason, the exponent $a$ at $T > 284$ K should decrease with an increase in the temperature. Thus, an increase in the exponent $a$ with the temperature at $T < T_d = 284$ K owing to a decrease in the potential barrier is replaced by its decrease with an increase in the temperature at $T > T_d = 284$ K. This results in the appearance of a maximum in the temperature dependence of the exponent $a$. This picture corresponds to the experimental data.